\documentstyle[aps]{revtex}
\newcommand{\ds}{\displaystyle}
\newcommand{\dsf}{\ds\frac}

\newcommand{\beq}{\begin{equation}}
\newcommand{\eeq}{\end{equation}}

\large

\begin{document}
\large

\begin{center}
\Large\bf

   STABLE THERMOMAGNETIC WAVES IN HARD SUPERCONDUCTORS

\vskip 0.1cm
{\normalsize\bf N.A.Taylanov}\\
\vskip 0.1cm
{\large\em Institute of Applied Physics, Department of Theoretical Physics,
National University of Uzbekistan, Tashkent, Uzbekistan
e-mail: taylanov@iaph.tkt.uz}
\end{center}
\begin{center}
\bf Abstract
\end{center}

\begin{center}
\mbox{\parbox{14cm}{\small
The problem of the stability of a nonlinear thermomagnetic wave with
respect to small thermal and electromagnetic perturbations in hard
superconductors was studied. It is shown that spatially bounded solutions
may correspond only to the perturbations decaying with time, which implies
stability of the nonlinear thermomagnetic wave.
}}
\end{center}
\vskip 0.5cm

Previously [1], it was demonstrated that, depending on the surface
conditions, stationary nonlinear thermomagnetic waves of two types,
$E$ and $H$, may exist in a superconductor. However, the problem of
stability of the nonlinear waves with respect to small thermal and
electromagnetic perturbations in superconductors is insufficiently studied.
We will consider the stability of a nonlinear thermomagnetic wave with
respect to small thermal and electromagnetic perturbations in a
superconductor occurring in the critical state. It will be demonstrated
that spatially bounded solutions may correspond only to the perturbations
decaying with time.
The evolution of the thermal ($T$) and electromagnetic ($E$ and $H$)
perturbations in a superconductor is described by a nonlinear thermal
conductivity equation [2]
\beq
\nu(T)\dsf{dT}{dt}=\Delta [\kappa (T)\Delta T]+\vec j\vec E,
\eeq
in combination with the Maxwell equations
\beq
rot\vec E=-\dsf{1}{c}\dsf{d\vec H}{dt}, \qquad rot{\vec H}=\dsf{4\pi}{c}\vec j
\eeq
and the critical state equation
\beq
\vec j=\vec j_{c}(T,\vec H)+\vec j_{r}(\vec E).
\eeq
where $\nu=\nu(T)$ and $\kappa=\kappa(T)$ are the heat capacity and
thermal conductivity coefficients, respectively; $\vec j_c$ is the critical
current density; and $\vec j_r$ is the resistive current density.
Let us consider a plane semiinfinite sample  $(x>0)$, exposed to an
external magnetic field $\vec H=(0, 0, H_{e})$, increasing with the time at a
constant rate $\dsf{d\vec H}{dt}=0$. According to the Maxwell equation (2),
the sample will feature a vortex electric field $\vec E=(0, E_e, 0)$
parallel to the current density vector: $\vec j$: $\vec E\parallel \vec j$.

In order to describe the $j_{c}(T,H)$ function, we will use the Bean-London
equation of the critical state $(\dsf{dj_c}{dH}=0)$ [3]. According to
this model, the $j_c(T)$ function can be presented as
$j(T)=j_{c}(T_0)\left[1-\dsf{T-T_{0}}{T_{c}-T_{0}}\right]$: where $j_{0}=j_{c}(T_{0})$
is the equilibrium current density, $T_c$ is the critical temperature,
and $T_0$ is the cooler temperature.
The characteristic form of  $j_{r}(E)$ in the region of sufficiently
strong electric fields $(E>E_f)$ can be approximated by a piecewise
linear function $j_r\approx\sigma_f E$. In the region of small field
strengths $E<E_f$ the $j_{r}(E)$ function exhibits a nonlinear character
related to a thermoactivated flux creep [4].
For an automodel solution of the type $\xi=x-vt$ describing a wave
propagating at a constant velocity $v$ along the $x$ axis, the system of
Eqs.( 1 )-(3) acquires the following form:
\beq
- v[N(T)-N(T_0)]=\kappa\dsf{dT}{d\xi}-\frac{c^2}{4\pi v}E^2,
\eeq
\beq
\dsf{dE}{d\xi}=-\dsf{4\pi v}{c^2}j,
\eeq
\beq
E=\dsf{v}{c}H.
\eeq

where $N(T)=\int\limits_0^T \nu(T)dT$.

    Excluding variables $T$ and $H$ with the aid of expressions (4) and (6)
and taking into account the boundary condition $E(z\to -\infty)=E_e$,
we arrive at a differential equation describing the distribution $E(z)$
\beq
\dsf{d^2 E}{dz^2}+\beta\tau\dsf{dE}{dz}+\dsf{4\pi v^2}{c^2E_\kappa}
\left[N(T)-N(T_0)\right]-\dsf{E^2}{2E_\kappa}=0
\eeq

Here $z=\dsf{\xi}{L}$ and $\beta=\dsf{vt_\kappa}{L}$ are dimensionless parameters,
$L=\dsf{cH_e}{4\pi j_0}$  the magnetic field penetration depth into the
superconductor,
$\tau=\dsf{D_t}{D_m}$ is the ratio of the coefficients of thermal
$D_{t}=\dsf{\kappa}{\nu}$ and magnetic
$D_{m}=\dsf{c^2}{4\pi\sigma_{f}}$ diffusion,
$t_\kappa=\dsf{\nu L^2}{\kappa}$ is the thermal diffusion time, and
$E_{\kappa}=\dsf{\kappa}{aL^2}$ is a constant parameter.

In the approximation of a weakly heated superconductor $(T-T_0)<<T_0$,
the heat capacity $\nu$ and thermal conductivity $\kappa$ coefficients
are weakly dependent on the temperature profile. As is well known
(see, e.g., [4]), the magnetic flux variations in a hard superconductor
occur at a much greater rate as compared to those of the heat transfer,
so that $\tau<<1$ or $D_t<<D_m$. In this approximation, we may neglect
the terms related to dissipative effects in Eq. (7), after which a solution
to this equation can be presented in the following form [5] :
\beq
E(z)=\dsf{E_1}{2}\left[1-th\dsf{\beta\tau}{2} (z-z_0)\right].
\eeq
Relationship (8) describes the profile of a thermomagnetic shock wave
propagating into the superconductor.
In order to study the stability of a nonlinear wave with respect to
small perturbations, it is convenient to write a solution to Eqs. (1)-(3)
in the following form:
\beq
T(z,t)={T}(z)+\delta T(z,t) \exp\left[\dsf{\lambda t}{t_\kappa}\right],
E(z,t)={E}(z)+\delta E(z,t) \exp\left[\dsf{\lambda t}{t_\kappa}\right],
\eeq

where ${T}(z)$ and ${E}(z)$ are the stationary solutions and
$\delta T, \delta E$  are small perturbations. Substituting expressions
(9) into Eqs. (1)- (3), assuming $\delta T$, $\delta E<<{T},{E}$
in the limit of $\tau<<1$ (this corresponds to a "fast" instability,
$\lambda>>\dsf{\nu v^2}{\kappa}$ [4]), we obtain an equation for
determining the eigenvalues of $\lambda$ [6]

\beq
\dsf{d^2\epsilon}{dz^2}+[\dsf{2}{ch^2}-\Lambda]\epsilon=0
\eeq

Using the substitution of variables of the type

$\xi=thy$,
$1-\xi=2s$,
and $\epsilon(\xi)=\left(1-\xi^2\right)^{\frac{\Omega}{2}}\Phi(\xi)$, Eq. (10)
can be reduced to a standart hypergeometric equation
\beq
s(1-s)\dsf{d^2\Phi}{ds^2}-[\Omega+1-2s(\Omega+1)]\dsf{d\Phi}{ds}-
[\Omega(\Omega-1)-p(p-1)]\Phi=0
\eeq
Here, the even and odd integrals can be presented as
\beq
\Phi_{1}=F(\epsilon-1, \epsilon+2, \epsilon+1, s),
\eeq
\beq
\Phi_{2}=s^{-\epsilon}F(-1, 2, 1-\epsilon, s),
\eeq
where F is the hypergeometric function [7]. The function $"$ must be finite
at the singular point $s=1$. The $\epsilon$ values for which $\Phi_{1}$
is finite evidently correspond to a discrete spectrum:
$\epsilon_{i}=0,1$ or $\lambda_{i}=-\tau^{-1},0.$
The function  $\Phi_{2}$ is finite only for $\epsilon=0$.
Any solution in the form of a running wave is char-
acterized by translational symmetry, which implies that a "perturbed"
stationary profile $E(z)$ determined by Eq. (8) corresponds to the
eigenvalue of the ground state $\lambda_{0}=0$. Differentiating Eq.
(7) with respect to $z$, one may readily see that $\dsf{dE_0}{dz}$
is an eigenfunction corresponding to $\lambda_0=0$. Indeed, the
perturbation $\delta E=\dsf{dE_0}{dz}$ essentially represents a
small wave displacement. Thus, we may suggest that tha function
$\dsf{dE_0}{dz}$ exponentially tends to zero for $z\rightarrow+\infty$
and the corresponding eigenvalue is zero. Therefore, the problem cannot
possess positive eigenvalues and $Re\lambda_{i}<0$. This result implies
that the wave is stable with respect to relatively small thermal
$\delta T$ and electromagnetic  $\delta E$ fluctuations.
The analysis of the second linearly-independent solution leads to the
same conclusion.

\end{document}